\documentstyle[12pt]{article}
\renewcommand{\thepage}{\arabic{page}}
\newcommand{\nc}{\newcommand}
\nc{\beq}{\begin{equation}} \nc{\eeq}{\end{equation}}
\nc{\beqa}{\begin{eqnarray}} \nc{\eeqa}{\end{eqnarray}}
\nc{\lsim}{\begin{array}{c}\,\sim\vspace{-21pt}\\< \end{array}}
\nc{\gsim}{\begin{array}{c}\sim\vspace{-21pt}\\> \end{array}}

\newcounter{mysection}
\newcounter{mysubsection}
\newcommand{\mysection}[1]{\stepcounter{mysection}\setcounter{equation}{0}
\setcounter{mysubsection}{0}\par\bigskip\noindent{\large\bf
\themysection .\ #1}\nopagebreak[4]\par\vskip .3cm}

\def\l:{\mathopen{:}\,}
\def\r:{\,\mathclose{:}}

\def\inbar{\,\vrule height1.5ex width.4pt depth0pt}
\font\cmss=cmss12 \font\cmsss=cmss12 at 7pt
\def\IZ{\relax\ifmmode\mathchoice
{\hbox{\cmss Z\kern-.4em Z}}{\hbox{\cmss Z\kern-.4em Z}}
{\lower.9pt\hbox{\cmsss Z\kern-.4em Z}}
{\lower1.2pt\hbox{\cmsss Z\kern-.4em Z}}\else{\cmss Z\kern-.4em
Z}\fi}
\def\IB{\relax{\rm I\kern-.18em B}}
\def\IC{{\relax\hbox{$\inbar\kern-.3em{\rm C}$}}}
\def\ID{\relax{\rm I\kern-.18em D}}
\def\IE{\relax{\rm I\kern-.18em E}}
\def\IF{\relax{\rm I\kern-.18em F}}
\def\IG{\relax\hbox{$\inbar\kern-.3em{\rm G}$}}
\def\IP{\relax{\rm I\kern-.18em P}}

\begin{document}

\begin{titlepage}

{\hbox to\hsize{hep-th/9804149 \hfill }}
{\hbox to\hsize{April 1998 \hfill TIFR-TH/98-10}}
{\hbox to\hsize{\hfill Fermilab-Pub-98/122-T}}
\bigskip

\begin{center}

\vspace{.5cm}

\bigskip

\bigskip

\bigskip

{\Large \bf  Three Brane Action and The 
Correspondence Between  N=4 Yang Mills Theory
and Anti De Sitter  Space }

\bigskip
{\bf Sumit R. Das}$^{\bf a}$  and {\bf Sandip P. Trivedi}$^{\bf b}$ \\

\bigskip

\bigskip

$^{\bf a}${ \small \it Tata Institute For Fundamental Research\\
 Homi Bhabha Road,  Mumbai  400 005, India \\}

\bigskip

$^{\bf b}${ \small \it Fermi National Accelerator Laboratory\\
  P.O.Box 500\\
 Batavia, IL 60510, USA\\}

\bigskip

\bigskip

{\bf Abstract}
\end{center}

Recently, a  relation between $N=4$ Super Yang Mills in $3+1$ dimensions and 
supergravity in an $AdS_5$ background  has been proposed. 
In this paper we explore the idea that the correspondence between 
operators in the Yang Mills theory and modes of the supergravity theory
can be obtained by using the D3 brane action. Specifically,
we consider two form gauge  fields for this purpose.  
The supergravity analysis predicts that the operator which corresponds to this 
mode has dimension six. We show that this is indeed the leading operator
in the three brane Dirac-Born-Infeld and Wess-Zumino action
which couples to this mode.  It is important in the analysis that the 
brane action is expanded around the anti de-Sitter background. Also,  the
Wess-Zumino term plays a crucial role in cancelling a lower dimension
operator which appears in the the Dirac-Born-Infeld action.

\end{titlepage}

\renewcommand{\thepage}{\arabic{page}}

\setcounter{page}{1}

\mysection{Introduction and Summary }

One of the interesting outcomes of recent progress in string theory
has been the relationship between gauge theories and gravity,
particularly in the context of black holes.
In particular, classical scattering of various
fields from non-dilatonic black holes like extremal three branes are
reproduced by correlators of the gauge theories living on the brane
worldvolume \cite{klebanovold}. Noting the fact that the near-horizon
geometry of such black holes is in fact a five dimensional anti-de
Sitter (AdS) space, Maldacena has conjectured that the large N limit of
a conformally invariant $d$- dimensional Yang Mills theory in fact
{\em contains} supergravity (and IIB superstring theory) 
in $(d+1)$ dimensional AdS space \cite{Malda}. This has led to
some progress in understanding the strong coupling behavior of these
gauge theories in the large-N limit.

The conjecture was further investigated in \cite{GKP} and
\cite{Witten1} where a concrete prescription was given for relating
observables in the supergravity and the Yang Mills theories. The idea
in \cite{GKP}, \cite{Witten1} was to consider $AdS$ space together
with a boundary. The dependence of the supergravity action on the
boundary values of fields then yields the required generating
functional from which Greens functions can be calculated.

In this paper we will be concerned with $3+1$ dimensional $SU(N)$ Yang
Mills theories with $N=4$ supersymmetry. The proposal of \cite{Malda}
relates this theory to ten- dimensional Type IIB supergravity
compactified on $ AdS_5 \times S_5$. Using the recipe outlined above,
several two point correlation functions were calculated in
\cite{GKP} and \cite{Witten1} and recently some three point functions
have been computed in \cite{mathur}. For some special operators, which
are either chiral or protected from renormalization effects for other
reasons agreement was found between the anomalous dimensions as
calculated in the Yang mills and the supergravity theories. For other
operators this led to a determination of the anomalous dimensions in
the large $N$ limit.

A large class of such operators are marginal or relevant in the
Yang-Mills theory. It is important to examine the relation between higher 
dimensional operators and supergravity modes as well.
 Some higher dimensional
operators have been studied in \cite{GHKK}. These may have two
different roles in the supergravity context. For some supergravity
modes, like the fixed scalar, these are responsible for the leading
order absorption by the black 3-brane and related to the correlator in
the AdS space itself. For other modes, like the dilaton, they are
responsible for corrections to the leading result and probe the
3-brane metric beyond the AdS throat \cite{GHKK}.

In this note we continue the study of this set of ideas by focussing
on another supergravity mode: the two index NS-NS antisymmetric tensor
field with a polarization parallel to some directions of the brane
\footnote{ Another mode is obtained by interchanging the roles of the
NS-NS and R-R two form fields, our results apply to this case as
well.}.  The propagation of this mode in AdS space was investigated in
\cite{KRN}, where, after incorporating the mixing with the R-R two
form field it's mass was determined.  More recently in \cite{Raja} the
cros-section for scattering this field off a extremal black hole was
calculated. These studies show that the scattering of the two form
field (with a polarization parallel to the brane) should be suppressed
for small energies.  Correspondingly the operator in the Yang Mills
theory to which it couples should have a total dimension of six.
Determination of this operator will be the goal of this paper.  We
should mention that this operator has been identified in
\cite{Ferrara}, \cite{AF}, from considerations of superconformal
symmetry \footnote{We thank the authors of \cite{Ferrara},\cite{AF}
for correspondence in this regard.}.

So far, in the literature, the principle behind a precise
correspondence between various modes in the supergravity theory and
operators in the Yang Mills theory has not been spelt out. In this
note we explore the idea that the operators of the Yang Mills theory
can be obtained by expanding an action consisting of the (non-abelian)
Dirac-Born-Infeld (DBI) and Wess-Zumino (WZ) terms.  We will see that,
for the antisymmetric tensor, the correct Yang Mills operator can be
identified in this way, but {\em only if} the action is expanded about
AdS spacetime and the accompanying five form field strength
background. In many ways this is the natural expectation.  The
supergravity mode being considered is a perturbation about $AdS$
space.  Thus one expects that to consistently couple it, the DBI
action should also be expanded about the $AdS$ background.  Related
points have been recently made in \cite{dealwis}. The conformal
symmetry of the three brane action has been studied in \cite{KT}.

It should be emphasized that here the DBI plus WZ action will be used
to identify the correct operators in the AdS-Yang Mills
correspondence. Whether higher order corrections to the correlator in
the full 3-brane geometry require the gauge field dynamics to be
governed by the DBI-WZ action remains to be seen. Some evidence in
favor of this has been presented in \cite{GHKK}.
 
One noteworthy feature about our analysis is that the WZ term
plays an important role in it.  The leading operator obtained from the
DBI action has engineering dimension of four.  But the
coupling to this operator is cancelled by a contribution coming from
the Wess-Zumino term. Thus the leading operator obtained from the
whole action, which couples to this supergravity mode, has dimension
six at tree level.  As was mentioned above, the supergravity analysis
shows that this must in fact be its total dimension.  We learn in this
way that the operator studied here does not acquire any anomalous
dimension in the large N limit. This is also true of the dimension
eight operator studied in \cite{GHKK}. 

One more point about our discussion below needs to be mentioned.  In
expanding the action we will need to decide where the DBI-WZ action
lives in AdS space.  As was mentioned above, the basic idea in the
discussion of \cite{GKP} and \cite{Witten1} is that the boundary
values for the supergravity modes act as sources for the Yang Mills
field operators.  This suggests the DBI-WZ action should be expanded
around the boundary of AdS space.  In fact, as we will see here, this
yields a consistent answer. We do not, however, take this to mean that
there are D3-branes physically located at the boundary.

Clearly, this analysis needs to be extended for other modes as well.
For one class of Yang Mills operators the coupling to the supergravity
modes has been written down from superconformal symmetry
considerations in \cite{Ferrara}.  It will be interesting to see if
these and the couplings to all the other supergravity modes can be
obtained by expanding the DBI-WZ action.  We hope to report more fully
on these questions in the future.  Let us mention one final point.
Our discussion in this note suggests that in the conformally
non-invariant cases, the supergravity theory should correspond to a
Yang Mills theory not in flat space-time but in the supergravity
background instead.

\mysection{The Supergravity Analysis}
A system of $N$ parallel D3- branes is described by an extremal
black hole geometry with a metric:
\beq
\label{metric}
ds^2=H^{-1/2}(dx^i)^2 + H^{1/2} (dx^a)^2,
\eeq
with
\beq
\label{defH}
H=1+{R^4 \over r^4}.
\eeq 
Here $x^{i}, i=0, \cdots 3$, refer to the four  coordinates parallel to the 
brane world volume, $x^a$ to the coordinates transverse to the brane,  
$r^2= (x^a)^2$ is the transverse coordinate distance away from the branes,
and $R^4=4 \pi g_s N (\alpha^{'})^2$, where $g_s$ is the string coupling,
related to the Yang-Mills coupling $g_{YM}$ by $g_s = g^2_{YM}$. 
In the near horizon region, $r \ll R$, this metric reduces to that 
of $AdS_5 \times S^5$:
\beq
\label{metricads}
ds^2={r^2 \over R^2} (dx^{i})^2 + {R^2 \over r^2} (dx^a)^2.
\eeq
we see that the $S_5$ has a radius $R$ and the geometry is smoothly varying
if $g_{YM}^2 N$ is large. 
The dilaton is constant in this background, while the self dual five form field 
strength in the near horizon region is given by:
\beq
\label{fourform}
F_{0123r}={r^3  \over R^4}.
\eeq

Following \cite{KRN} and \cite{Raja} we now consider the 
propagation of the NS-NS  and R-R two form fields
in the AdS background. 
The NS-NS and R-R gauge potentials will be denoted by $B_{\mu \nu}$ and
$A_{\mu \nu}$ and the corresponding fields strengths by $H_{\mu \nu \rho}$
and $F_{\mu \nu \rho}$ respectively.
The equations governing the propagation of these modes are \cite{JHS}:
\beqa
\label{propeq}
\nabla^{\mu}H_{\mu \nu \rho}&=&({2 \over 3}) F_{\nu \rho \kappa \tau \sigma} 
                                          F^{\kappa \tau \sigma} \nonumber \\
\nabla^{\mu}F_{\mu \nu \rho} &=&-({2 \over 3}) F_{\nu \rho \kappa 
        \tau \sigma}  H^{\kappa \tau \sigma}. 
\eeqa
 
We see that in the presence of a non-zero five-form  field strength the 
$B_{\mu \nu}$ and $A_{\mu \nu}$ fields mix with each other. 

Here we only consider an  s-wave mode with the NS-NS two form being polarised
along the brane directions.  For simplicity,  the only component of 
$B_{\mu \nu}$ that is non-zero will be chosen to be  $B_{12}$. 
Eq. (\ref{propeq}) can now be used to solve for $F_{\mu \nu \rho}$ 
in terms of $B_{1 2}$ and gives \footnote{In our conventions $F_{\mu \nu \rho}=
\partial_{\mu} A_{\nu \rho} + \partial_{\nu} A_{\rho \mu} + \partial_{\rho}
A_{\mu \nu} $. } :
\beq
\label{valF}
F_{0r3}= -4 ~ F_{r3012} ~g^{11} g^{22} ~B_{12},
\eeq
with all the other components of the three form RR field strength being
zero. In the subsequent discussion we consider a perturbation with 
energy $\omega$, with a resulting time dependence $e^{-i \omega t}$.  
>From eq. (\ref{propeq}), eq. (\ref{valF}),  we then get an  
equation for $B_{12}$:
\beq
\label{decB}
{1 \over {\sqrt -g}} {1 \over g^{11}} {1 \over g^{22}} 
 \partial_r({\sqrt -g} g^{rr} g^{11} g^{22} \partial_r B_{12}) - w^2 g^{00} 
                               B_{12} =
-16 F_{0123r}^2 g^{00} g^{rr} g^{33} g^{11} g^{22} B_{12}.
\eeq

Substituting for $F_{0123r}$, from eq. (\ref{fourform})  now gives:
\beq
\label{finalprop}
{1 \over {\sqrt -g}} {1 \over g^{11}} {1 \over g^{22}} 
 \partial_r({\sqrt -g} g^{rr} g^{11} g^{22} \partial_r B_{12}) - w^2 g^{00} 
                               B_{12} = {16 \over R^2} B_{12}.
\eeq
Thus we see that the mode corresponds to a field with mass $m= {4 \over R}$. 
In the correspondence between  Supergravity  and Yang Mills theory
the region, $r \gg R$ is particularly relevant. 
 In this region the second term on the 
left hand side of eq. (\ref{finalprop} ) can be neglected,
giving rise to two solutions with, $B_{12} \sim r^{\pm 4}$. 
Of these the case,
\beq
\label{solB}
B_{12}  \sim  c ~r^4,
\eeq
can be extended to a non-singular solution in the small $r$ region, it 
will be the relevant one for the subsequent discussion. 

We can also  solve for the R-R two form $A$, corresponding to 
eq. (\ref{solB}). From  eq. (\ref{valF}), in the $r \gg R$ region
it is given by: 
\beq
\label{solA}
A_{03} = - B_{12}
\eeq
with all other components being zero.  

Now that our analysis of the supergravity mode is complete we can use 
the prescription of \cite{GKP}, \cite{Witten1} to calculate the 
anomalous dimension of the Yang Mills operator that corresponds to it.
The general formula relating the anomalous dimension, $\Delta$, to the mass
for a $p$ form is:
\beq
\label{relmassdim}
(\Delta -p)(\Delta+p-4)=m^2,
\eeq
where the mass is measured in units of $R$.
In this case, $m^2=16$, eq. (\ref{finalprop}), setting in addition $p=2$, 
gives:
\beq
\label{valdelta}
\Delta = 6.
\eeq
Thus the operator in the Yang Mills theory corresponding to the mode discussed 
here has a total dimension of $6$ in the large N limit. 
In the next section  we turn to determining this operator.

Before doing so though, let us pause to make contact with the general analysis
in \cite{KRN}. Here we have focussed on an S-wave mode of the two index gauge 
field. In \cite{KRN} the propogation of all the higher Kaluza Klein harmonics
arising from the two form field (and in fact all the other supergravity modes)
were  analyzed as well. The propogation equation for the two form,  
eq. (2.56) in \cite{KRN}, 
was elegantly factorised giving rise to two families,  eq. (2.62), 
and eq. (2.64) of \cite{KRN} (also shown in FIG 3 as the two $a_{\mu \nu}$ 
modes). The S-wave mode discussed in this paper is the $k=0$ member of the 
second family, eq. (2.64). The  first family, eq. (2.63), only involves $l=1$ 
and higher angular momentum modes. In fact the $l=1$ mode of this family
(which transforms like a $6$ of $SU(4)$ and is the mode shown in Fig 3 with a 
circle) is discussed in \cite{Ferrara} where it was 
identified with a dimension three operators in the Yang Mills theory.

\mysection{Identifying the Operator in the Yang Mills Theory}
So far in the literature on this subject, a precise procedure for
identifying operators in the Yang Mills theory, that correspond to a
particular perturbation mode in the supergravity theory, has not been
given.  The supergravity analysis determines the dimension of the
operator. In some cases supersymmetry determines the operator,
e.g. the conserved currents in \cite{Ferrara}. However, 
even in these cases the overall normalisation is not determined {\em a priori}.
For example,  the relevant   power of the string 
coupling in the normalisation cannot be determined in this manner.  
 This point becomes
clear when one tries to extract absorption cross-sections from the
Yang-Mills theory. The leading power of energy in the Yang-Mills
calculation is determined by the total dimension of the operator,
but the power of string coupling depends on further details of the
operator, as is clear from the analysis of \cite{das}.

Here, we will explore the idea that these operators can be obtained by
expanding an action containing a DBI and WZ term
\footnote{Strictly speaking we are proposing to expand the action
which governs the dynamics of D3-branes. If additional terms besides
the DBI and WZ terms are present in such an action, as has been
suggested in
\cite{AT}, one would expect to keep them as well.    
We note here  that the additional terms discussed in \cite{AT} are of 
dimension $12$ and higher; such high dimension operators do not alter our 
conclusions.}. 
  This method for identifying operators has been used
earlier in \cite{fixed} for five dimensional black holes and in
\cite{GHKK} for the 3-brane, where the action was expanded about flat
space-time and shown to give consistent results.  The present
discussion will have two important new features.  First, as we will
discuss, it will be crucial to expand the action about AdS space, rather
than flat space, to
identify the operator correctly.  In many ways, this is the natural
thing to do. The supergravity modes we are interested in are
perturbations about AdS space. Thus to couple them consistently one
also expects to expand the brane action about the AdS background.
Secondly, our analysis will involve the WZ term in an important
way.  In particular a cancellation, between two contributions , which
arise from the DBI and WZ terms respectively, to the coupling of a
dimension four operator, will be important in identifying the leading
operator.  We will see that this procedure yields a consistent result.
In fact, the leading operator which couples to the mode of section 2
has dimension six.  The supergravity analysis showed that it's total
dimension is six as well. Thus, we find that the operator does not
acquire any anomalous dimensions in the large N limit.

It is worth drawing attention to one other aspect of the analysis at
the outset.  The action we use can be identified with the world volume
theory for a set of D3- brane probes.  In the discussion below we will
take these branes to be placed at the boundary of AdS space,  at
$r \gg R$.  This is in line with the discussion of \cite{GKP},
\cite{Witten1}.  In these references, the idea was to compute correlation
functions by, roughly speaking, regarding boundary values of the
supergravity fields as sources for the Yang Mills theory .  This
implies that the Yang Mills operators should also be identified by
working at the boundary of AdS space.

The action for a D3-brane has been studied in \cite{Ceder},
\cite{APS}, and is given by \footnote{We have set the dilaton
$e^{\phi}=1$ and chosen units for the string tension so that the
coefficient in front of the DBI term is unity.}:
\beq
\label{braneaction}
S=-\int d^4 \xi {\sqrt { -det(G_{ij} +{\cal F}_{ij} )} } + \int ({\hat
C_{(4)}} + {\cal F} \wedge {\hat A} + {\hat C_{(0)}} 
{\cal F} \wedge {\cal F} ).
\eeq
The two terms above correspond to the DBI action and the
WZ term respectively.  It is worth defining the various terms
above carefully.  $G_{ij}$ refers to the induced world-volume metric,
obtained as the pull-back of the spacetime metric. Similarly,
\beq
\label{defcalF}
{\cal F}_{ij}= F_{ij} - {\hat  B_{ij}},
\eeq
where $F_{ij}$ stands for the gauge field on the D3-brane and ${\hat
B_{ij}}$ is the pullback of the NS-NS two form potential. In the W-Z
term ${\hat C_{(4)}}$,  ${\hat A }$ and ${\hat C_{(0)}}$ refer to the
 pullback of the R-R four form,  two form and zero form fields  respectively. 
 Strictly speaking we are
interested here in the action for $N$ D3 branes. It has been suggested
in \cite{tseytlin} that this can be obtained by appropriately
symmetrising the terms obtained from the single brane action.  To
begin with  we will work with the abelian single brane action.  The color
factors will be introduced by appropriate symmetrisation towards the
end.  Some of our conclusions do not depend on the details of this
symmetrisation procedure.

In the following discussion it is useful to distinguish between three
kinds of indices: $\xi^i, i=0, \cdots 3$, refer to the world-volume
coordinates and are purely bosonic.  $Z^M$, refers to ten dimensional
superspace coordinates; $M$ can stand for bosonic coordinates denoted by
$m=1, \cdots 10$, or for fermionic coordinates denoted by $\mu$.
Finally, we denote frame indices by $A$; $A$ can stand for bosonic
tangent vectors, denoted by $a=1, \cdots 10$, or for spinor tangent
vectors denoted by $\alpha$.  In this note we will be interested in
expanding this action about AdS space to linear order in the
perturbation, eq. (\ref{solA}), eq. (\ref{solB}).  Following,
\cite{APS} we will choose a static gauge, where $X^m= \xi^m, m=0 ,
\cdots 3 $.  The remaining $X^m, m=4 \cdots 9$ will be denoted by $\phi^m$. 
In addition the kappa symmetry of eq. (\ref{braneaction}) will be used
to set half the fermionic coordinates to zero \footnote{Our
conventions for spinors and Dirac matrices are the same as those in
\cite{APS}. Namely, the $\Gamma$ matrices are $ 32 \times 32$ real matrices
satisfying, $\{\Gamma^{\mu}, \Gamma^{\nu} \}= 2 \eta^{\mu \nu}$, with
 $\eta = (-1, 1 , \cdots 1)$.} \cite{APS}.  The remaining $16$ component Majorana-Weyl 
fermion will then be denoted by $\lambda$.

With this notation in hand the induced metric is given by
\beq
\label{pullmetric}
G_{ij}={\partial Z^M \over \partial {\xi^i} } {\partial Z^N \over
\partial {\xi^j} }~ E_M^A ~E_N^B \eta_{AB}.
\eeq
with $E_M^A$ being given by:
\beqa
\label{flatvielbein}
E_m^a  &=& {r \over R} \eta^a_m,  m=0 ,\cdots  3 \nonumber \\
              &=& {R \over r}  \eta^a_m,   m=4 ,\cdots  9 \nonumber \\
E_{\mu}^a &=&  ({\bar \lambda} \Gamma^a)_{\mu} \nonumber \\
E_m^{\alpha} &=& 0 \nonumber \\
E_{\mu}^{\alpha} &=&\delta_{\mu}^{\alpha} \nonumber.
\eeqa
>From eq. (\ref{pullmetric}), (\ref{flatvielbein}) it then follows that
the induced metric is given by \footnote{ To avoid confusion let us
note that all indices on Gamma matrices here are frame indices. }:
\beqa
\label{finalmetric}
G_{ij} & =& {r^2 \over R^2} \eta_{ij} +  {R^2 \over r^2}
{\partial \phi^m \over \partial \xi^i}  {\partial \phi^n \over \partial \xi^j} 
\eta_{mn}  \nonumber \\
&& - \left ( {\bar \lambda}~ (\Gamma_i {r \over R} +
\Gamma_m  {\partial \phi^m \over \partial \xi^i} {R \over r} )~ 
{\partial \lambda \over \partial \xi^j} + i\leftrightarrow j  \right ) +
\sum_{a=0}^9 {\bar \lambda} \Gamma^a {\partial \lambda \over \partial
\xi^i} ~ {\bar \lambda} \Gamma_a {\partial \lambda \over \partial
\xi^j}.
\eeqa

Similarly, the pullback of the rank-2 tensor is defined as
\beq
\label{bdef}
B_{ij} = B_{AB} \partial_i Z^M \partial_j Z^N E_M^A E_N^B
\eeq
and the superspace components of $B_{AB}$ are given in \cite{Ceder}. 
Using these we get 
\beqa
\label{finalF}
{\cal F}_{ij}&=&F_{ij}- \left ( {\bar \lambda}~ (\Gamma_i {r \over R}
+ \Gamma_m {\partial \phi^m \over \partial \xi^i} {R \over r})~
{\partial \lambda \over \partial \xi^j} - i
\leftrightarrow j \right ) -~B_{ij} \nonumber \\ &&+ {R \over r}~\left
(B_{ik}~ {\bar \lambda} \Gamma^k \partial_j \lambda -i\leftrightarrow
j \right ) - {R^2 \over r^2}~({\bar \lambda}
\Gamma^k {\partial_i \lambda}~ {\bar \lambda} \Gamma^l {\partial_j \lambda}
~ B_{kl} )
\eeqa

Adding eq. (\ref{finalmetric}) and (\ref{finalF}) we get that the DBI
action is given by :
\beq
\label{BIe}
S_{DBI}=-\int d^4  \xi ~({r \over R} )^4~  {\sqrt  {-det(\eta_{ij} +M_{ij} )} }
\eeq
where 
\beqa
\label{defM}
M_{ij} &=& {R^4 \over r^4} \partial_i \phi^m \partial_j \phi^n
\eta_{nm} + {R^2 \over r^2} F_{ij} -2 {\bar \lambda } [{R \over r} \Gamma_i+ 
{R^3 \over r^3} \Gamma_m \partial_i\phi^m]\partial_j\lambda \nonumber
\\ 
&&+ {R^2 \over r^2} \sum_{a=0}^{9}({\bar \lambda} \Gamma^a \partial_i
\lambda ) ({\bar \lambda} \Gamma_a \partial_j \lambda ) - 
{R^2 \over r^2} B_{ij} \nonumber \\ 
&&+{R^3 \over r^3} [ B_{ik} {\bar \lambda} \Gamma^k \partial_j \lambda -
B_{jk} {\bar \lambda} \Gamma^k
\partial_i \lambda ] -{R^4 \over r^4} B_{kl}~({\bar \lambda} \Gamma^k
\partial_i \lambda)~ ({\bar \lambda} \Gamma^l \partial_j\lambda).
\eeqa
                    
This gives rise to  the following  terms in the DBI action  linearly
dependent on the NS-NS two form :
\beqa
\label{expandedBI}
\int  d^4  ~\xi L_{BI}&=& \int d^4 \xi   \left [ 
- {1\over 2} B_{ij} F^{ji} + {r \over R} ~{\bar \lambda }~
\Gamma_i \partial_j \lambda 
                       B^{ji} \right ] -{R \over r}  
\left [ {\bar \lambda} \Gamma^m F_{mi} \partial_j
\lambda B^{ij} \right ]  \nonumber \\ 
 &&- \left [ {1 \over 2} {R^4 \over r^4} F_{im} F^{mj} 
F_{jk} B^{ki} \right] + {1\over 2}{R \over r}B_{ij}F^{ji}{\bar \lambda}~
\Gamma^l \partial_l \lambda \nonumber \\
&& -{1\over 8}{R^4 \over r^4}F_{lm}F^{ml}F_{ij}B^{ij} + \cdots.
\eeqa
The ellipses above refer to operators of dimension eight and higher
that occur in the expansion.  Also, for the sake of brevity we have
regrouped terms so that the indices above take values in ten
dimensions and are raised and lowered by the flat space metric.  Thus
for example, $F_{im}$ refers to the ten dimensional field strength
\footnote{The last two terms in (\ref{expandedBI}) were not included
in the first version of this paper, we regret this error}.

The action described above is for a single D3-brane. While a definitive
action is not known for $N$ branes, we may adopt the symmetrized trace
prescription of \cite{tseytlin}, so that the various quantities
above have to be replaced by matrices and traced over. The first
term in (\ref{expandedBI}) then receives a contribution only from
the $U(1)$ piece of $U(N)$ and is subdominant in the large N limit.
Thus the lowest dimension operator which
couples to $B_{ij}$ in eq. (\ref{expandedBI}) has dimension $4$ and is
given by the first term in eq. (\ref{expandedBI}).  The subsequent
operators listed in eq.  (\ref{expandedBI}) all have dimension $6$.

The contribution from the WZ term arises from the wedge product of
${\cal F}$ and ${\hat A}$ in eq. (\ref{braneaction}), which can be
expanded as:
\beq 
\label{wzexpanded}
L_{WZ}= {1 \over 4} {\cal F}_{ij} {\hat A_{kl}} \epsilon^{ijkl}.
\eeq
We remind the reader that in eq. (\ref{wzexpanded}) ${\hat A_{kl}}$ 
stands for the pullback of the R-R two form.
This is given by:
\beq
\label{defpullrr}
{\hat A_{ij}}= A_{ij} - {R \over r} \left [ {\bar \lambda} \Gamma^k 
\partial_j \lambda A_{ik} - i \leftrightarrow j \right ] + \cdots
\eeq
where the ellipses again denote higher dimensional operators. 
>From eq. (\ref{solA}) it follows that   in the general case, for  a perturbation
$B_{ij}$ polarised along the brane directions, the R-R two form is given by
\footnote{Here all indices are being summed using the flat space metric. 
Our  conventions for the  $\epsilon$ symbol are   chosen so  that ,
$\epsilon^{0123}=+1$. }:  
\beq 
\label{solAgeneral}
A_{kl}= {1\over 2}~\epsilon_{klij} ~ B^{ij}.
\eeq
Substituting eq. (\ref{defpullrr}) and eq. (\ref{solAgeneral}) 
in  eq. (\ref{wzexpanded})  then leads to :
\beq
\label{finalwz}
L_{WZ} = -{r \over R} {\bar \lambda} \Gamma^i \partial_j 
                                 \lambda B^{ji} 
                  -{R \over r}  {\bar \lambda}~  \Gamma^m \partial_j 
                                 \lambda~F_{im}~B^{ij} 
+{1 \over 2} {R \over r} F_{ij}B^{ij} {\bar \lambda} \Gamma^l 
\partial_l \lambda.
\eeq
In obtaining
eq. (\ref{finalwz}) we have used eq. (\ref{finalF}) for ${\cal F}$ and
expanded keeping only terms linear in $B_{ij}$.
As in eq. (\ref{expandedBI}), we have regrouped terms and expressed them
in a notation where all indices take values in ten dimensions (the  indices are raised and lowered by the flat space metric). 
Finally, for reasons mentioned above, we have omitted a term in 
eq. (\ref{finalwz}) which goes like $B_{ij}F^{ij}$.   

On adding eq. (\ref{expandedBI}) and (\ref{finalwz}) we now see
that the dimension four operators that couple to $B_{ij}$ do indeed
cancel between the two equations \footnote{ As was mentioned above,
we have so far been suppressing color indices.  For the dimension four
operator involved here there is a unique color singlet that can be
formed ; the cancellation then follows in the full Non- Abelian theory
as well. }.  Furthermore the two types of dimension six operators
which involves the fermionic fields also cancel between 
(\ref{expandedBI}) and (\ref{finalwz}). 
Thus the leading operators are of dimension six and of
the form:
\beq
\label{finaldimsix}
{\cal O}_6 = -({R  \over r})^4 ~ \left [  {1 \over 2}~ F_{jm}F^{mk}F_{ki} 
B^{ij} + {1 \over 8} F_{lm} F^{ml} F_{ij} B^{ij} \right ] .
\eeq
While we have suppressed the color indices here, the operators in eq.
(\ref{finaldimsix}) should be understood as being symmetrised in color
space.  Eq. (\ref{finaldimsix}) is the main result of this paper.

A few comments are in order at this stage.  

Our starting point was the assumption that the coupling between
various supergravity modes and operators in the Yang Mills theory can
be obtained by expanding the action consisting of the DBI and W-Z terms
about AdS space.  The consistent answer obtained above provides
evidence in support of this assumption.  Note in particular that the
analysis above probed terms in the action of dimension six and
involved the W-Z term in a non-trivial way.

It was crucial in the above analysis to expand the action about AdS
space.  What would have happened if we had
expanded about flat space? In this case, without any five -form field
strength, the supergravity equations, eq. (2.5), would not have coupled
the NS-NS and R-R two form gauge potentials together and would have
lead to two massless modes.  Correspondingly, on expanding the brane
action for the coupling to the NS-NS two form, one would have only got
a contribution of the form of eq. (3.8) from the DBI action with
no contribution from the WZ term.  Thus the leading operator in the
Yang Mills theory would have had dimension four which is  different 
from what we got above.

It will be interesting to study this set of ideas further for the
other supergravity modes as well.  We have looked at the S-wave mode
for the two form fields here. As was mentioned above, the propagation of 
higher partial waves was studied in \cite{KRN}. One would like to 
determine the corresponding Yang Mills operators as well \footnote{
Here we have been implicitly assuming that the action is expanded about 
a definite position in the $S_5$. 
In correctly determine the operators for the 
higher partial waves we will probably need to average this position over
the whole of  $S_5$.}.
  For one class of supergravity modes (including an $l=1$ mode of the two 
form field) the coupling to the Yang Mills theory has been written down in
\cite{Ferrara} from superconformal considerations. 
It is worth examining if these couplings can be obtained by expanding
the DBI plus WZ action.  

\mysection{Note Added in revised version :} 

While this revised version was being written \cite{FLZ} appeared,
where a large class of gauge theory operators corresponding to
supergravity modes, including the operator derived in this paper, have
been shown to follow from the proposal of this work. This paper has
also noted the omission of the second term in ${\cal O}_6$ in
our earlier version.

\mysection{Acknowledgement}

We would like to thank Joe Lykken for comments and especially Samir
 Mathur for sharing his insights and numerous
 discussions. S.R.D. would like to thank Center for Theoretical
 Physics, M.I.T and Physics Department of Stanford University for
 hospitality during the course of this work. The research of S.T.  is
 supported by the Fermi National Accelerator Laboratory, which is
 operated by Universities Research Association, Inc., under contract
 no. DE-AC02-76CHO3000.

\nc{\ib}[3]{ {\em ibid. }{\bf #1} (19#2) #3}
\nc{\np}[3]{ {\em Nucl.\ Phys. }{\bf #1} (19#2) #3}
\nc{\pl}[3]{ {\em Phys.\ Lett. }{\bf #1} (19#2) #3}
\nc{\pr}[3]{ {\em Phys.\ Rev. }{\bf #1} (19#2) #3}
\nc{\prep}[3]{ {\em Phys.\ Rep. }{\bf #1} (19#2) #3}
\nc{\prl}[3]{ {\em Phys.\ Rev.\ Lett. }{\bf #1} (19#2) #3}

\end{document}